\documentclass{ethpaper}
\pdfoutput=1
\usepackage{ifpdf}
\usepackage{graphicx}
\usepackage{amssymb}
\usepackage{subfigure}
\begin{document}
\begin{titlepage}
\ethnote{}
\title{A study of high-energy proton induced damage\\
in Cerium Fluoride  in comparison with measurements\\in Lead Tungstate calorimeter crystals}
\begin{Authlist}
G. Dissertori, P.~Lecomte, D.~Luckey, F.~Nessi-Tedaldi, F.~Pauss
\Instfoot{eth}{Institute for Particle Physics, ETH Zurich, 8093 Zurich, Switzerland}
Th.~Otto, S.~Roesler, Ch.~Urscheler
\Instfoot{psi}{Institute for Particle Physics, ETH Zurich, 8093 Zurich, Switzerland}
\end{Authlist}
\maketitle
\begin{abstract}
  A CeF$_3$ crystal produced during early R\&D studies for 
  calorimetry at the CERN Large Hadron Collider
  was exposed to a 24\,GeV/c proton fluence $\Phi_p=(2.78 \pm
  0.20) \times 10^{13}\;\mathrm{cm^{-2}}$ and, after one year of
  measurements tracking its recovery, to a fluence $\Phi_p=(2.12 \pm 0.15) \times
  10^{14}\;\mathrm{cm^{-2}}$.  Results on proton-induced damage to
  the crystal and its spontaneous recovery after both irradiations
  are presented here, along with some
  new, complementary data on proton-damage in Lead Tungstate.
  A comparison with FLUKA Monte Carlo simulation results is performed and a
  qualitative understanding of high-energy damage mechanism is
  attempted.
\end{abstract}

\vspace{7cm}
\conference{submitted to Elsevier for publication in Nucl. Instr. and Meth. in Phys. Research A}

\end{titlepage}
\section{Introduction}
\label{s-INT}
The Large Hadron Collider (LHC) at CERN is expected to undergo a
substantial upgrade in luminosity after the exploitation of its
physics potential, to allow further exploration of the high energy
frontier in particle physics. The higher luminosity will strengthen
the requirements on performance of most detector components. With data
taking extended over several years under an increased luminosity,
detectors will be exposed to even larger ionising radiation and hadron
fluences than at the LHC, and they will need to be 
upgraded as well. While the harvest of LHC proton-proton collision data is
just starting, studies are already being performed on a LHC upgrade
(superLHC) where calorimetry will have to perform adequately in a
radiation environment and hadron fluences an order of magnitude more
severe than at the LHC.  It will thus be important to have results at
hand on calorimetric materials able to withstand the anticipated radiation
levels and particle fluences before making decisions on detector
upgrades.

Concerning hadron effects on crystals used for calorimetry, the
present study extends to Cerium Fluoride (CeF$_3$) and complements our
earlier work performed on Lead Tungstate
(PbWO$_4$)~\cite{r-LTNIM,r-pionNIM,r-LYNIM}.  In our earlier
investigations we have shown that high-energy
protons~\cite{r-LTNIM} and pions~\cite{r-pionNIM} cause a permanent,
cumulative loss of Light Transmission in PbWO$_4$,
while we observed no hadron-specific
change in scintillation emission~\cite{r-LYNIM}. 
The features of the observed damage hint at disorder that might be caused by 
fragments of heavy elements, Pb and W. Above a certain threshold, these 
can have a range up to 10 $\mu$m and energies up to $\sim$100 MeV,
corresponding to a stopping power nearly 10000 higher than the one of 
minimum-ionising particles. The associated local energy deposition
is capable of inducing the displacement of lattice atoms.

The qualitative understanding we gained of hadron damage in Lead
Tung\-state lead us to predict~\cite{r-CAL08} that such hadron-specific
damage contributions are absent in crystals, like Cerium Fluoride, consisting only of
elements with $Z < 71$, which is the experimentally observed threshold
for fission~\cite{r-THR}. Studies on Cerium Fluoride are expected to help at the same time
 in understanding  hadron damage to scintillating crystals and in
possibly providing a viable material for calorimetry in an extreme environment as the
superLHC will be.

\section{Cerium Fluoride}
Cerium Fluoride is a scintillating crystal whose luminescence
characteristics are known since the early studies by F.A. Kr\"oger and
J.~Bakker~\cite{r-KRO}, who measured its emission spectrum and light
decay time constants and understood the responsible transitions.  Its
properties as a scintillator were revealed by
D.~F.~Anderson~\cite{r-AND} and by W.~W.~Moses and
S.~E.~Derenzo~\cite{r-MOS}, who attracted attention to its
characteristics as a fast, bright and dense calorimetric medium for
high-energy physics and positron-emission tomography
applications.

Its density ($\rho=6.16\; {\mathrm{g/cm}}^3$), radiation length ($X_0 =
1.68$ cm), Moli\`ere radius ($R_M = 2.6$ cm), nuclear interaction length
($\lambda_I = 25.9$ cm) and refractive index ($n = 1.68$) make it a
competitive medium for compact calorimeters. Its emission is centred
at 340 nm, with decay time constants of 10 - 30 ns; it is insensitive
to temperature changes (${\mathrm{dLY/dT}}\; (20^o$C$)=0.08\%/^o$C) as well as bright (4 -
10\% of NaI($T\ell$)) and thus suitable for high-rate, high-precision
calorimetry~\cite{r-AND2}.

\begin{figure}[h]
\begin{center}
{\mbox{\includegraphics[width=12cm]{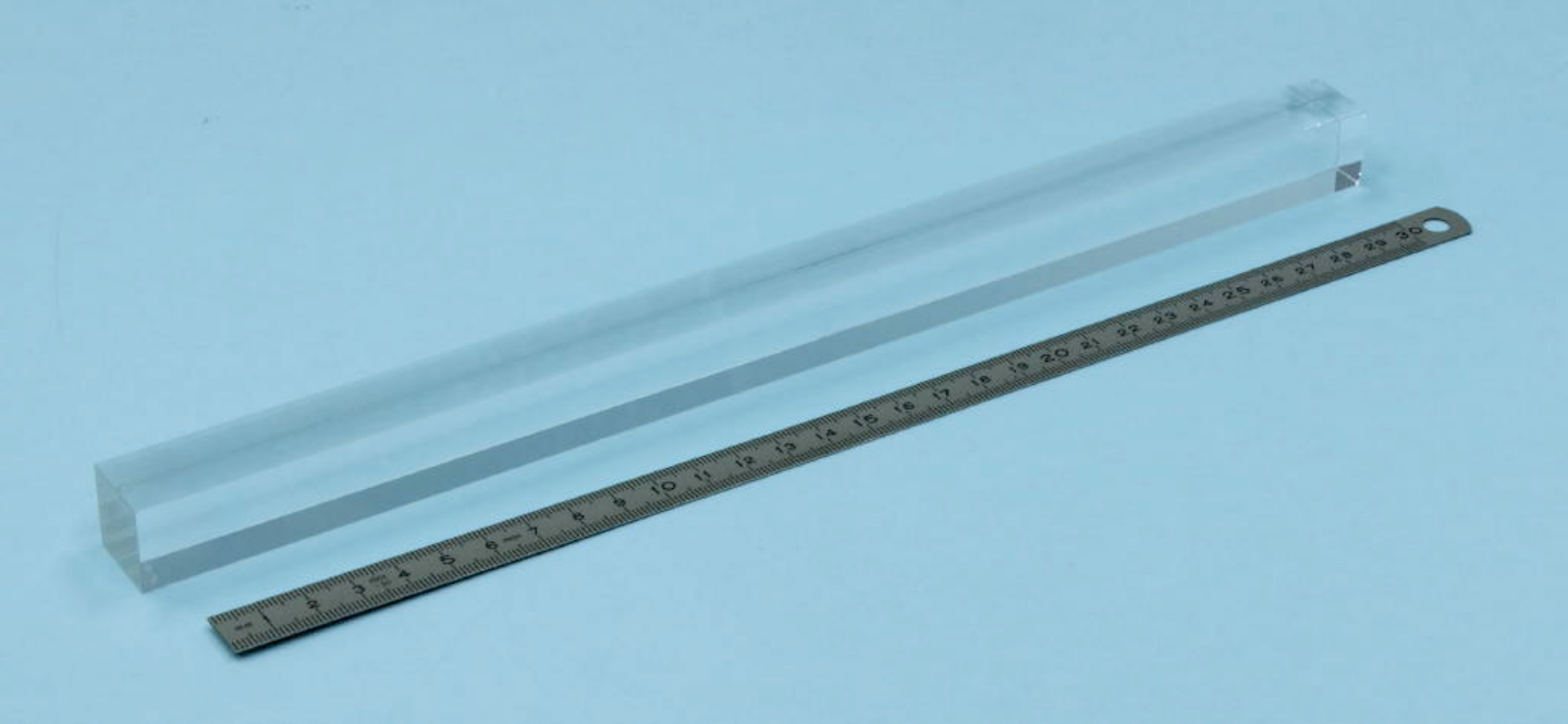}}}
\end{center}
\caption{A CeF$_3$ crystal, 30 cm long, produced by the
Shanghai Institute of Ceramics.}\label{f-PHOTO}
\end{figure}

In the nineties, this material was subject to an intense research
program, that established its scintillation characteristics, its
behavior in $\gamma$~\cite{r-KOB,r-CCC1} and MeV-neutron
irradiations~\cite{r-CHI} and the capability for crystal growers to
produce crystals of dimensions suitable for high-energy physics
applications.  It should be noted that Cerium is a readily available
rare earth, which would allow containing raw material costs, were a
mass production envisaged. Its melting point of $1430^o$C allows
applying well-known crystal growth technologies. The ability to grow
crystals beyond 30 cm length was demonstrated, as visible in
Fig.~\ref{f-PHOTO}, but for its use in a calorimeter, R\&D would
have to be resumed, since no commercial production of macroscopic
crystals presently exists, although the material is still used, e.g. in the
form of 10 $\mu$m nanoparticles, for neutron capture
cross-section measurements~\cite{r-STA}.

The tysonite structure is complicated, with each Cerium atom surrounded by
11 Fluorines in the Edshammar's polyhedron~\cite{r-HYD} and
the lower symmetry thus causes difficulties in the calculation of energy
levels and defect structures~\cite{r-MER}.
Cerium Fluoride is superionic~\cite{r-TRN}, the fluorine ions having
high mobility, which could account for part of its observed radiation
hardness. It is also paramagnetic, a characteristics which influences
the Faraday rotation in a magnetic field~\cite{r-XU}.  Fluorine ions
labelled $F_1$ in literature~\cite{r-F1} account for most of
the Cerium Fluoride conductivity and hence their mobility might help
repair defects. There is a vast literature on this and in particular
on doping~\cite{r-ROS} with Barium and Strontium, which have been
observed to increase the conductivity even further~\cite{r-PRI}.  The
composition which maximizes conductivity might maximize radiation
hardness as well, as hypothesised in ~\cite{r-SOR}, but establishing
this would require a dedicated R\&D.  It should also be
noticed that the Cerium Fluoride conductivity increases by a factor 4
between $20^o$C and $50^o$C temperature~\cite{r-SOR}, while its Light
Output remains unaffected.  Thus, for calorimetry applications in a
hostile radiation environment, maintaining the crystals at a higher
temperature could help minimising radiation damage. It is
also pointed out in ~\cite{r-SOR} that such crystals are the best fluoride superionics
for electro-chemical solid state devices such as fuel cells.

The performance of Cerium Fluoride was also studied with high-energy
particle beams in prototype crystal matrices~\cite{r-CEFTB}, in
particular since it was adopted as baseline calorimetric medium in the
CMS ~\cite{r-LOI} and L3P ~\cite{r-L3P} Letters of Intent.  Energy
resolutions of the order of $0.5\%$ for electron energies of 50 GeV
and higher were achieved, and it was observed how Cerium Fluoride
appeared to be the best material at that time for homogeneous
electromagnetic calorimetry at LHC and only the need for a very
compact calorimeter justified relegating it behind Lead Tungstate 
as the preferred material.

Cerium Fluoride was also considered for the ANKE spectrometer upgrade
at COSY and for medical imaging
applications~\cite{r-MO2}. In~\cite{r-NOV}, its energy response to
electromagnetic probes was extended down to a few MeV in photon energy
and a good time resolution, below 170 ps, was obtained using a time-of-flight
technique.

We have performed a test of hadron effects in Cerium Fluoride with the
expectation to yield a
better understanding of the whole hadron damage issue in
scintillating crystals, and to provide the community with a viable
solution for the hadron fluences expected during operation at
superLHC.
\section{The crystal}
\label{s-xtal}
For this study, we have used a Barium-doped $\mathrm{CeF}_3$ crystal from
Optovac~\cite{r-optovac}, which has parallelepipedic dimensions of $21
\times 16 \times 141\;\mathrm{mm}^3$ $(8.4 \;\mathrm{X_0})$.  Its
longitudinal Light Transmission (LT) before irradiation as a function
of wavelength is shown in Fig.~\ref{f-LT}.  One observes that the
smoothness of the transmission curve is interrupted by several dents
which are known to be due to the presence of $\mbox{Nd}^{3+}$
impurities~\cite{r-CCC2} and are of no further concern to the present
study.  One also observes, in the light of Fig. 2 in
Ref.~\cite{r-CCC2}, how the Barium doping translates into a
characteristic Light Transmission band edge which sits right above 300
nm, i.e. $\sim$ 15 nm higher compared to crystals grown with undoped
raw material.
\section{The irradiations}
\label{s-irrad}
The crystal was irradiated with 24\,GeV/c protons at the IRRAD1
facility\,\cite{r-IR1} in the T7 beam line of the CERN PS accelerator.
The first irradiation was performed beginning of November 2007, with a
flux $\phi_p= 1.16 \times 10^{12}$\,cm$^{-2}$h$^{-1}$. The proton
fluence reached was $\Phi_p=(2.78 \pm 0.20) \times
10^{13}\;\mathrm{cm^{-2}}$.  After one year of periodic measurements, where
its spontaneous recovery at room temperature, in the dark, was tracked, a second
irradiation was performed with a flux $\phi_p= 0.94 \times
10^{13}$\,cm$^{-2}$h$^{-1}$ and a similar measurement series was
performed.  The proton fluence reached with the second irradiation was
$\Phi_p=(2.12 \pm 0.15) \times 10^{14}\;\mathrm{cm^{-2}}$.  In both
cases, the irradiation procedure described in
\cite{r-LTNIM}, where all details can be found, was followed: the proton beam was
broadened to cover the whole crystal front face, and the fluence for
each irradiation was determined through the activation of an aluminium
foil covering the crystal front face.

 \section{Light Transmission measurements and results}
\begin{figure}[h]
\begin{center}
  \begin{tabular}[h]{cc}
{\mbox{\includegraphics[width=80mm]{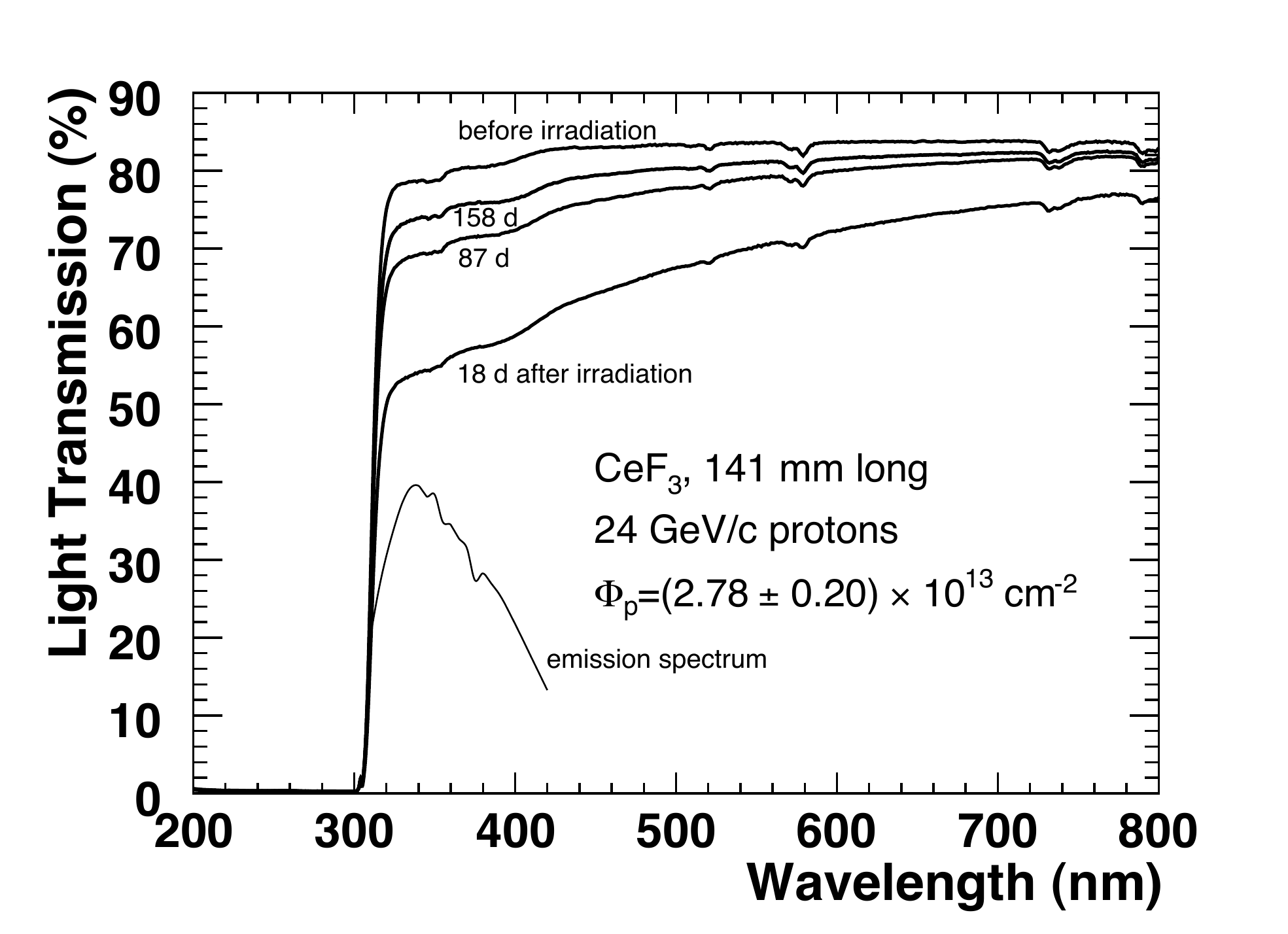}}} &
{\mbox{\includegraphics[width=80mm]{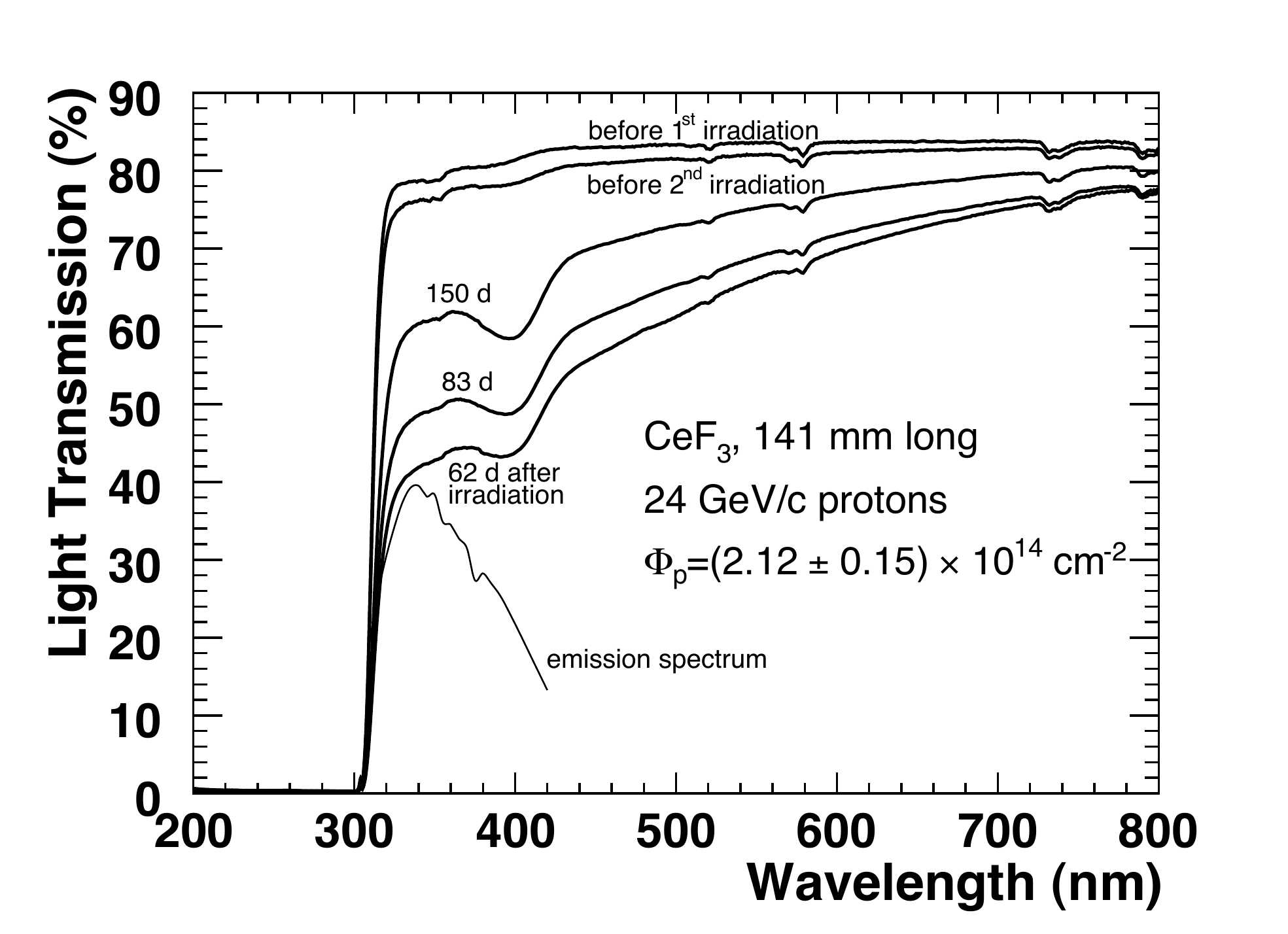}}}
   \end{tabular}
\end{center}
\caption{Transmission curves for Cerium Fluoride before and at various
  times after the first irradiation (left figure) and before and at various
  times after the second irradiation (rightfigure).\label{f-LT}}
\end{figure}

Longitudinal transmission curves at various intervals after proton irradiation are
represented in Fig.~\ref{f-LT}. The earliest ones were taken as soon
as it was possible to handle the crystal while keeping people's
exposure to radiation within regulatory safety limits, 18 days after
the first irradiation and 62 days after the second one.  From the LT
curves, it is evident that the damage reduces Light Transmission at
all wavelengths, while no transmission band-edge shift is observed
after irradiation. This observation is consistent with our qualitative
understanding~\cite{r-IEEE}, that the band-edge shift observed in Lead
Tungstate~\cite{r-LTNIM} and BGO~\cite{r-BGO} after hadron irradiation
must be due to disorder causing an Urbach-tail
behavior~\cite{r-URB,r-ITO}.  The absence of a band-edge shift in
Cerium Fluoride is consistent with the anticipated lack of heavy
fragments that can cause lattice disorder. The extreme steepness of
the band-edge, which is preserved throughout the proton irradiations,
is due to an allowed transition, as indicated in~\cite{r-SCH}.
 \label{s-LT}

The longitudinal Light Transmission was repeatedly measured over time, to collect
recovery data.  The damage is quantified through the induced
absorption coefficient as a function of light wavelength $\lambda$,
defined as:
\begin{equation}
\mu_{IND}(\lambda) = \frac{1}{\ell}\times \ln \frac{LT_0 (\lambda)}{LT (\lambda)}
\label{muDEF}
\end{equation}
where $LT_0\; (LT)$ is the Longitudinal Transmission value measured
before (after) irradiation through the length $\ell$ of the crystal.
\begin{figure}[h]
\begin{center}
{\mbox{\includegraphics[width=10cm]{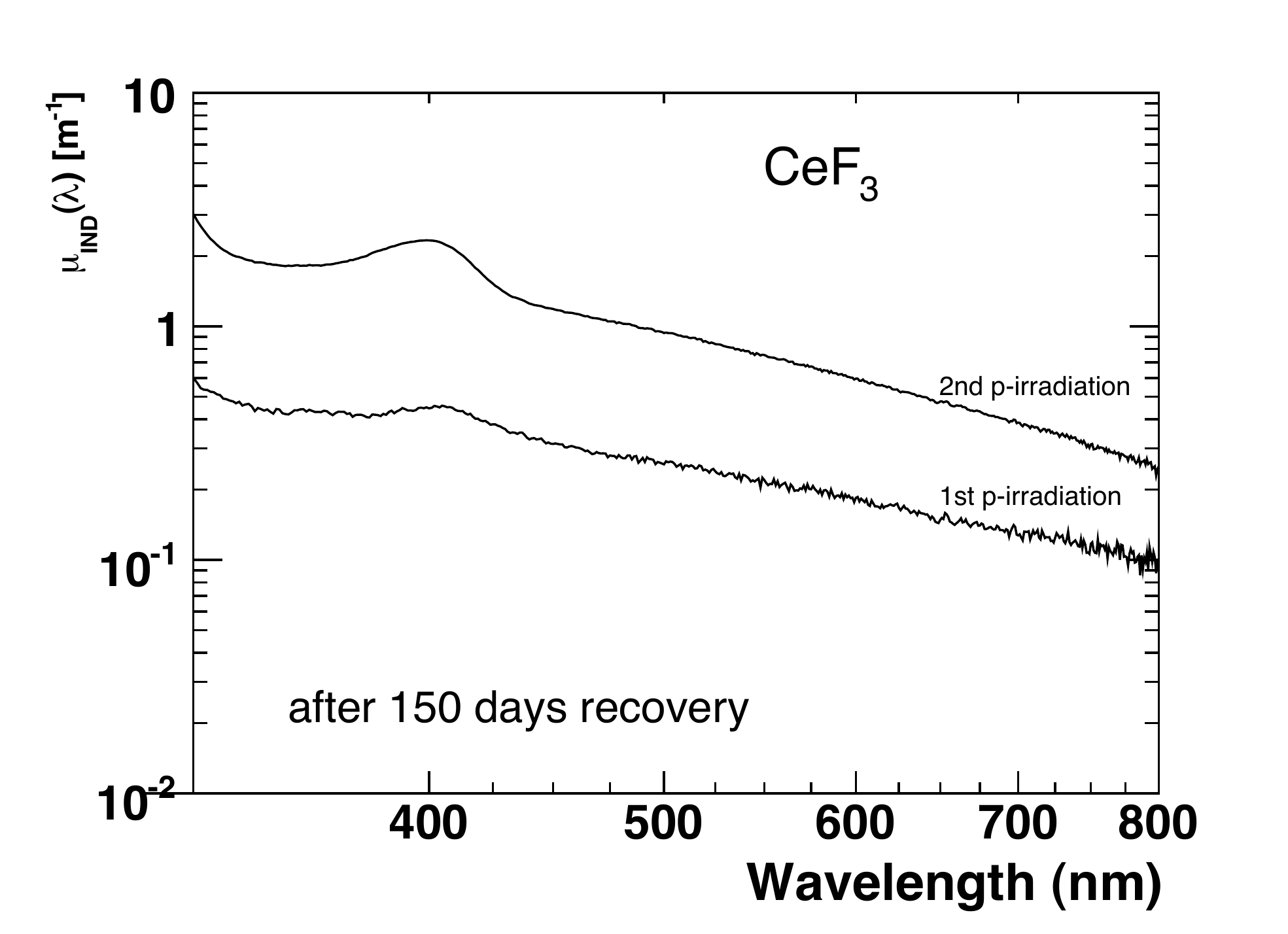}}}
\end{center}
\caption{Induced absorption coefficient as a function of wavelength in
Cerium Fluoride,
  150 days after irradiation after the first and after the
  second proton irradiation.\label{f-mu}}
\end{figure}

Figure~\ref{f-mu} shows the profile of induced absorption as a
function of wavelength, 150 days after each irradiation.  We notice
here the absence of the $\lambda^{-4}$ behavior we previously
observed~\cite{r-LTNIM} in Lead Tungstate. That behavior, peculiar to
Rayleigh scattering, is a qualitative indication of the presence of
very small regions of severe damage, as one expects to be caused by
highly ionising fragments from nuclei break-up.  The absence of a
Rayleigh-scattering behavior in Cerium Fluoride is a further
confirmation of our understanding.  We also observe the
presence of a yet unidentified absorption band, peaked around 400 nm,
which does not recover with time, but is of no further concern, because
it affects only a small fraction of the emitted light.
The absorption band amplitude scales in a way
which is consistent with a linear dependence on $\Phi_p$. The density of
centres $N$ multiplied by the oscillator strength $f$ calculated according
to~\cite{r-DEX} for the second irradiation is
\begin{equation}
N \times f \simeq 1.7 \times 10^{13} {\mbox{cm}}^{-3}.
\end{equation}
A FLUKA~\cite{r-fluka1,r-fluka2} simulation of the irradiation, which is
described in detail in section \ref{s-ACT}, yields an abundance $\rho
= 2.4\times 10^{-2} {\mbox{cm}}^{-3}$ of stable light nuclei (H and He)
per impinging proton. Even assuming an oscillator strength $f = 1$,
such an abundance does not account for the observed absorption band,
and thus allows to exclude defects caused by hydrogen and helium.
Typical $f$-values range from 1 to
$10^{-3}$ and therefore we hypothesise that the observed absorption
band might be linked to defects in the Cerium sub-lattice.

Furthermore, the dips due to Nd$^{3+}$ contamination mentioned in section
~\ref{s-xtal}
disappear when the induced absorption coefficients are evaluated, as
it is evident in the plots of Fig.~\ref{f-mu}.  This proves that
such dips are not influenced by radiation, nor are hidden absorption
bands present underneath them.

To examine damage recovery further, data analysis focused on the
changes in Longitudinal Transmission at a wavelength of 340 nm, which
corresponds to the peak emission of $\mbox{CeF}_3$ scintillation
light, and is thus the relevant quantity for calorimetry.
\begin{figure}[h]
\begin{center}
  \begin{tabular}[h]{cc}
{\mbox{\includegraphics[width=80mm]{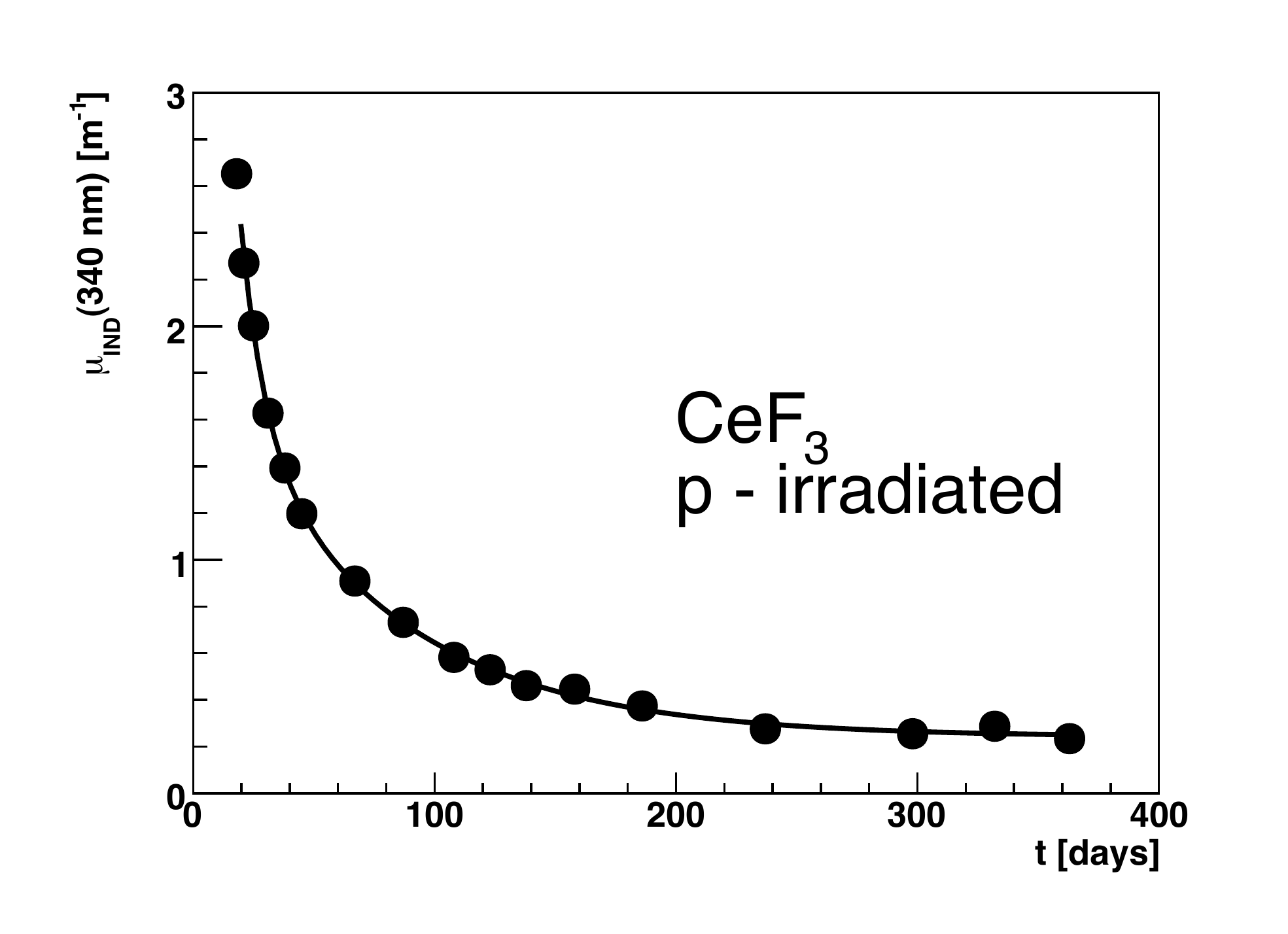}}}&
{\mbox{\includegraphics[width=80mm]{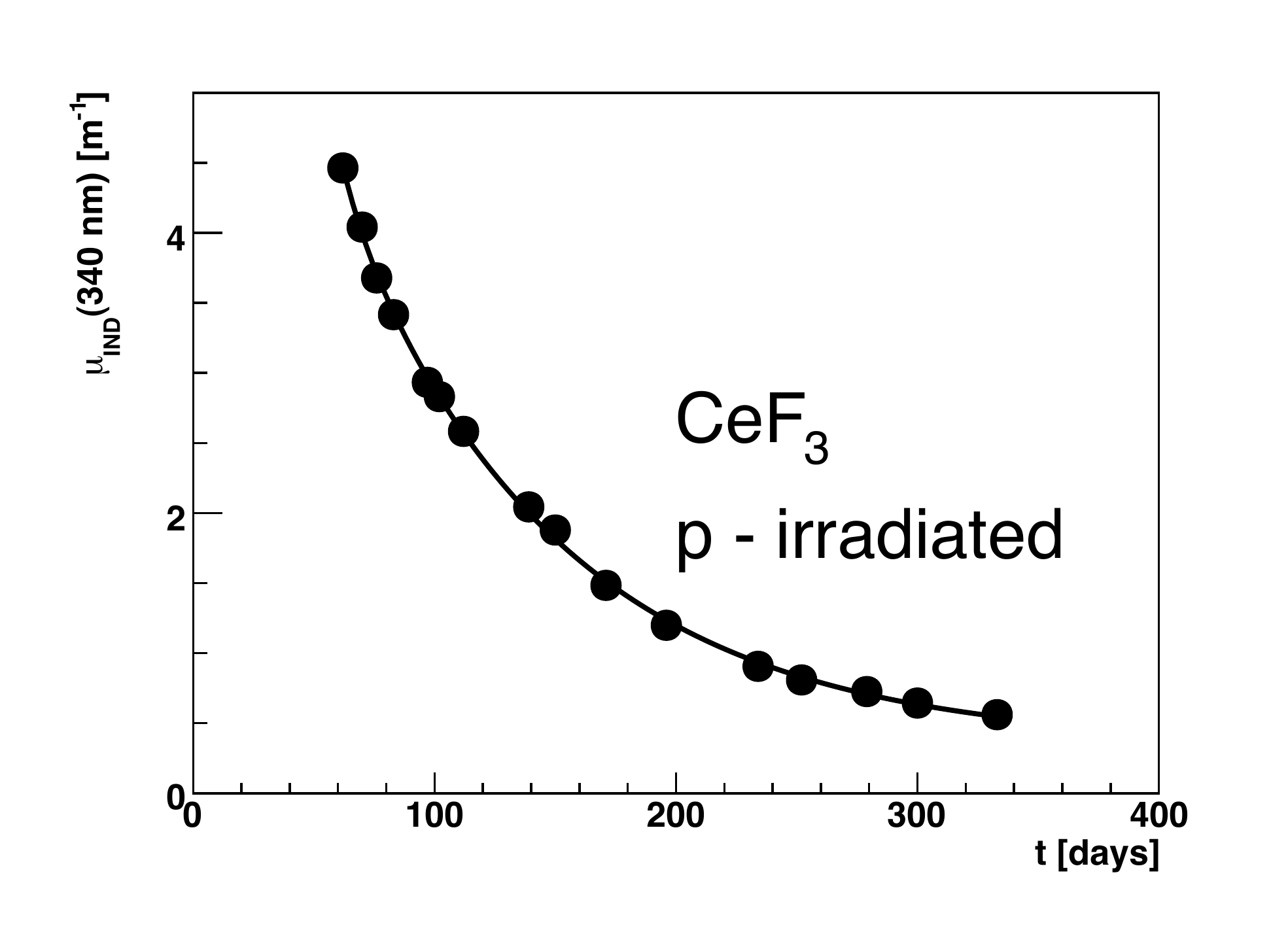}}}
   \end{tabular}
\end{center}
\caption{Recovery curves for Cerium Fluoride after the first (left) and
  after the second (right) proton irradiation\label{f-recLT}}
\end{figure}
The evolution of damage over time is shown in Fig.~\ref{f-recLT} for
the two irradiations, where $\mu_{IND}(340\; \mbox{nm})$ is plotted
over time.  The data, taken over one year, are well fitted by a sum of
a constant and two exponentials with time constants $\tau_i\;
(i=1,2)$:
\begin{equation}
\mu_{IND}^{ j}(340\; \mathrm{nm},t_{\rm{rec}}) = \sum_{i=1}^{2} A_i^je^{-t_{\rm{rec}}/\tau_i} + A_3^j
\label{e-Afit}
\end{equation}
\begin{figure}[t]
\begin{center}
{\mbox{\includegraphics[width=11cm]{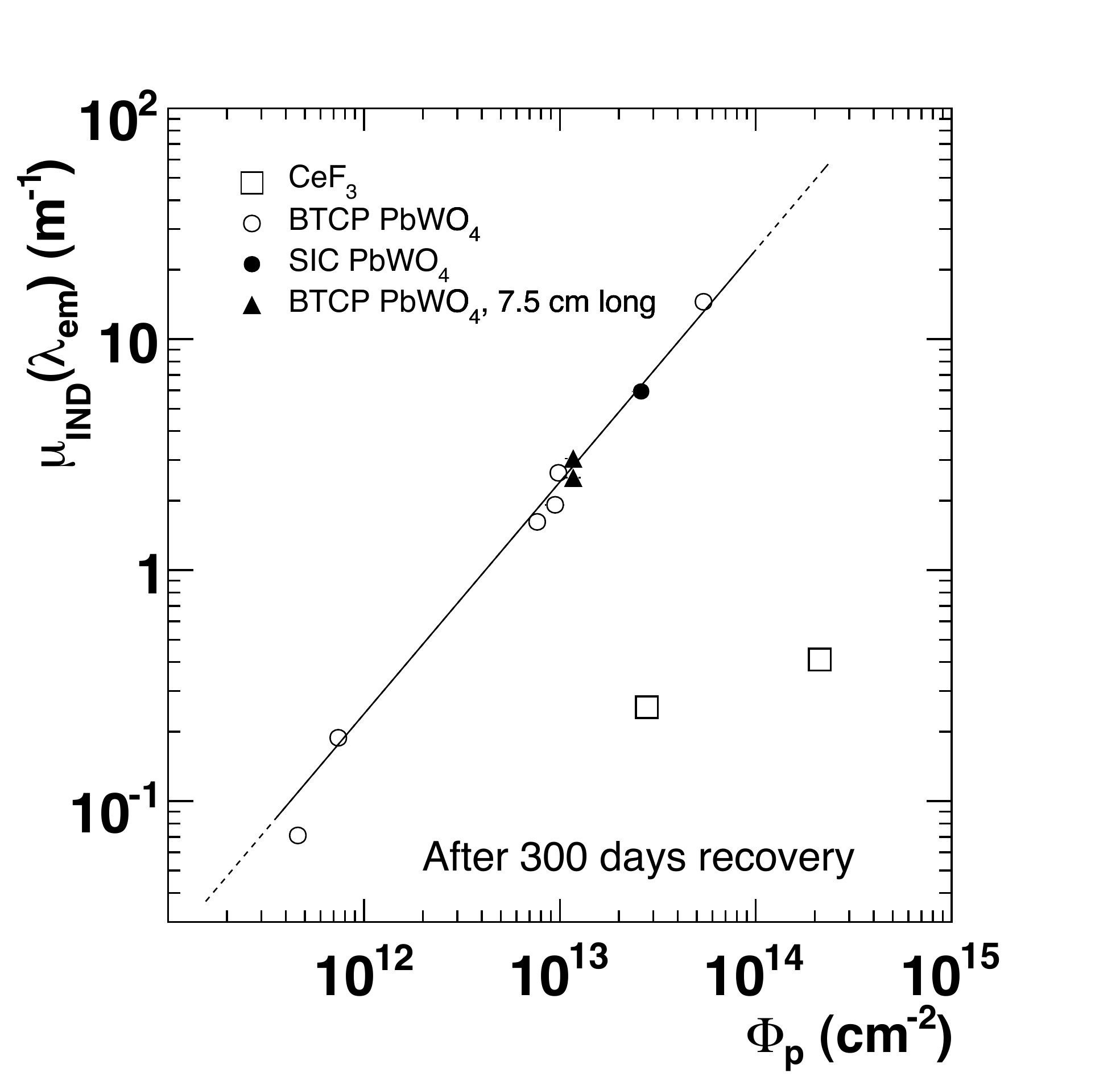}}}
\end{center}
\caption{Induced absorption coefficient versus proton fluence
for Cerium Fluoride and
for Lead Tungstate crystals of different length and origin.\label{f-LIN}}
\end{figure}
where $t_{\rm{rec}}$ is the time elapsed since the irradiation, while
$A_i^j \; (i=1,2)$ and $A_3^j$ are the amplitude fit parameters for
the irradiation $j\; (j=1,2)$.  Figure \ref{f-recLT} shows a fit where
the recovery time constants have been independently fitted for the two
irradiations. The time constants obtained are compatible, yielding values
$\tau_1 = 11\pm 2$ days and $\tau_2 = 70\pm 9$ days, with all the damage
recovering. Such a result is very important if one considers Cerium Fluoride for
superLHC calorimetry, in that one can expect to have a hadron damage
which remains - due to its recovery time characteristics - all the time at a
very small level compared to the cumulative damage amplitudes we
measured for Lead Tungstate~\cite {r-LTNIM}. In all this it should be
pointed out, however, that our measurements are not sensitive to
damage with a recovery time constant shorter than a few days, because
the proton-irradiated crystal was initially too radioactive for safe handling
during the first two weeks after irradiation, and thus no measurements were performed on it.

The dependence of damage on proton fluence is plotted in
Fig.~\ref{f-LIN}, in comparison with the values obtained
for Lead Tungstate. The
line therein is the fit from Fig.~15 in~\cite{r-LTNIM} to
$\mu_{IND}(420\;\mbox{nm})$ for Lead Tungstate, 150 days after
irradiation. The white circles are measurements of
$\mu_{IND}(420\;\mbox{nm})$ for the same Lead Tungstate crystals
studied in~\cite{r-LTNIM}, taken 300 days after
irradiation, showing how stable the long-term damage is in that
crystal. Since in~\cite{r-LTNIM} all crystals studied were produced by
the Bogoroditzk Techno-Chemical Plant (BTCP) in Russia, we have performed an
irradiation of a further crystal produced by a different supplier, the
Shanghai Institute of Ceramics (black circle in Fig.~\ref{f-LIN}). The
measurement for the SIC Lead Tungstate crystal is in perfect agreement
with those for BTCP crystals, proving how hadron damage is not
linked to fine details of doping, stoichiometry and related defects,
nor to growth technology, but it is rather due to the effects the
hadron cascade has on the bulk of the crystal.

It should be pointed out that all crystals tested in
Ref.~\cite{r-LTNIM} are 23 cm ($25.8\; X_0$) long, while the Cerium
Fluoride crystal studied here is only $14.1$ cm ($8.4\; X_0$) in
length.  In the same plot, we have thus also superposed the proton
damage measured in two Lead Tungstate crystals $7.5$ cm ($8.4\; X_0$)
long, studied in~\cite{r-pionNIM} after a proton irradiation where
they were placed one behind the other.  Also for these shorter
crystals $\mu_{IND}(420\;\mbox{nm})$ is well consistent with the
values measured for the longer crystals. This can be understood from
the star density profiles  for 20 - 24 GeV/c protons
in Fig.~3 of ref.~\cite{r-LTNIM}, which are
nearly flat over the length of the crystal, besides a small build-up
over the initial 5 cm, which is reflected in a slightly smaller damage
in one of the two crystals.  It thus appears justified to compare
damage measured in an $8.4\; X_0$ long Cerium Fluoride crystal with
the existing measurements for 23 cm long Lead Tungstate crystals.  The
induced absorption $\mu_{IND}(340\;\mbox{nm})$ at the peak of
scintillation emission measured in Cerium Fluoride 300 days after
irradiation is thus also plotted in Fig.~\ref{f-LIN}. As easily
understandable, a cumulative damage in Cerium Fluoride would be fitted
by a line parallel to the one for Lead Tungstate in this
doubly-logarithmic plot, which is not what we observe.

With the correlation of Fig.~\ref{f-LIN} extending over almost three
orders of magnitude in fluence, the damage
observed 150 days after irradiation is a factor 15 smaller in Cerium
Fluoride than in Lead Tungstate for a fluence $\Phi_p=2.78 \times
10^{13}\;\mathrm{cm^{-2}}$ and a factor 30 smaller for $\Phi_p=2.12
\times 10^{14}\;\mathrm{cm^{-2}}$.  The gap increases to a factor 25
and 124 respectively 300 days after irradiation, as also visible in
Fig.~\ref{f-LIN}.  However, one has to be aware of the fact that, were
the induced absorption expressed in units of inverse radiation length,
the gap would be reduced by a factor 1.5.

\section{Light Output measurements and results} 
\label{s-LY}
The relevant quantity for calorimeter operation is the Light Output
(LO), and thus, in the present study we have verified that with the
recovery of Light Transmission, also the Light Output is restored.
For this purpose, we have taken Light Output spectra using a
bialkaline 12-stage Photomultiplier (PM), and its anode charge was
digitised using a charge-integrating ADC as described
in~\cite{r-LYNIM}. To identify a scintillation signal well above the
background due to the intrinsic induced radioactivity after
proton irradiation, we have
triggered on cosmic muons traversing the crystal sideways, by means of
two plastic scintillators.
\begin{figure}[h]
\begin{center}
  \begin{tabular}[h]{cc}
{\mbox{\includegraphics[width=55mm,angle=90]{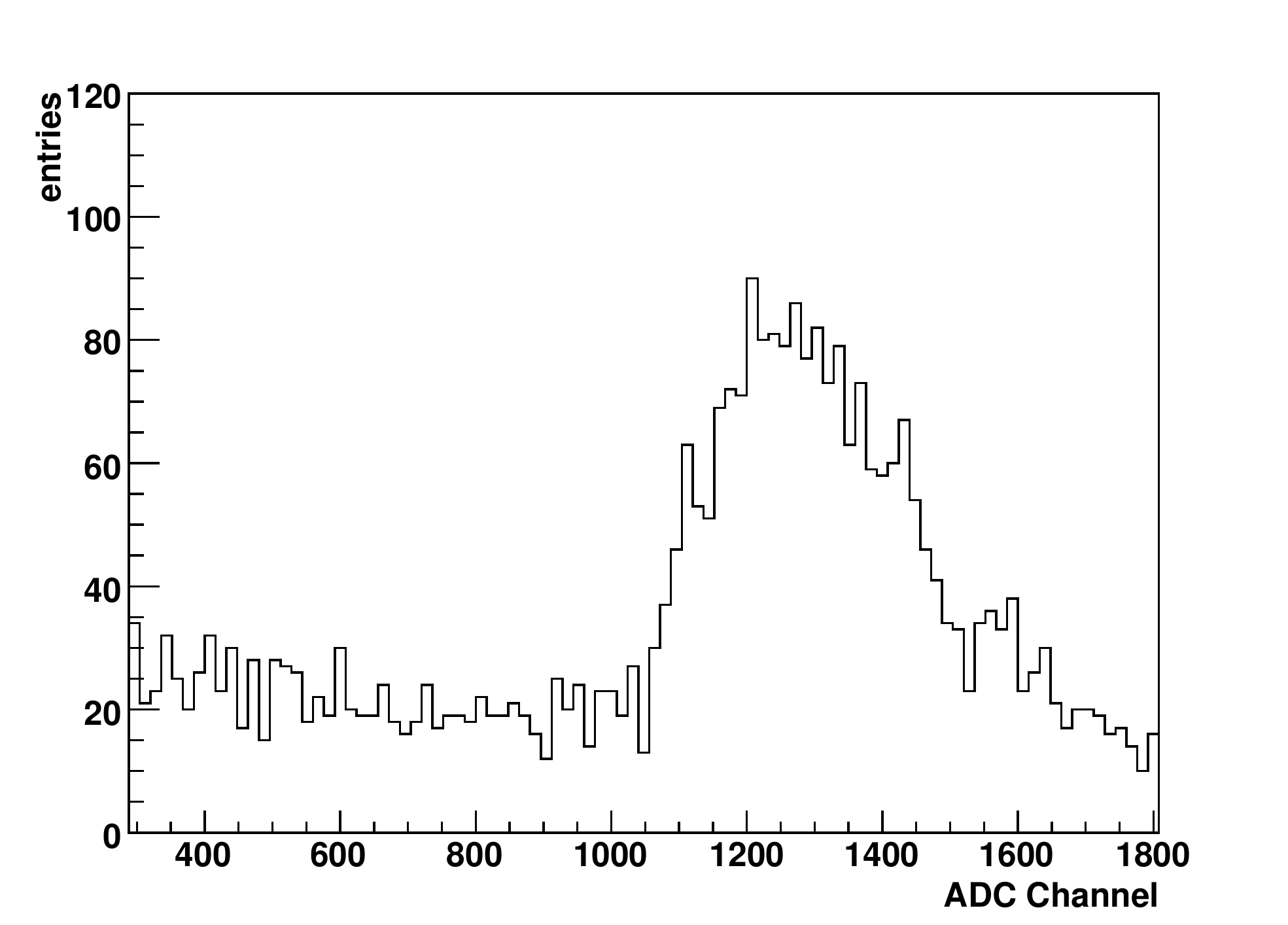}}}&
{\mbox{\includegraphics[width=55mm,angle=90]{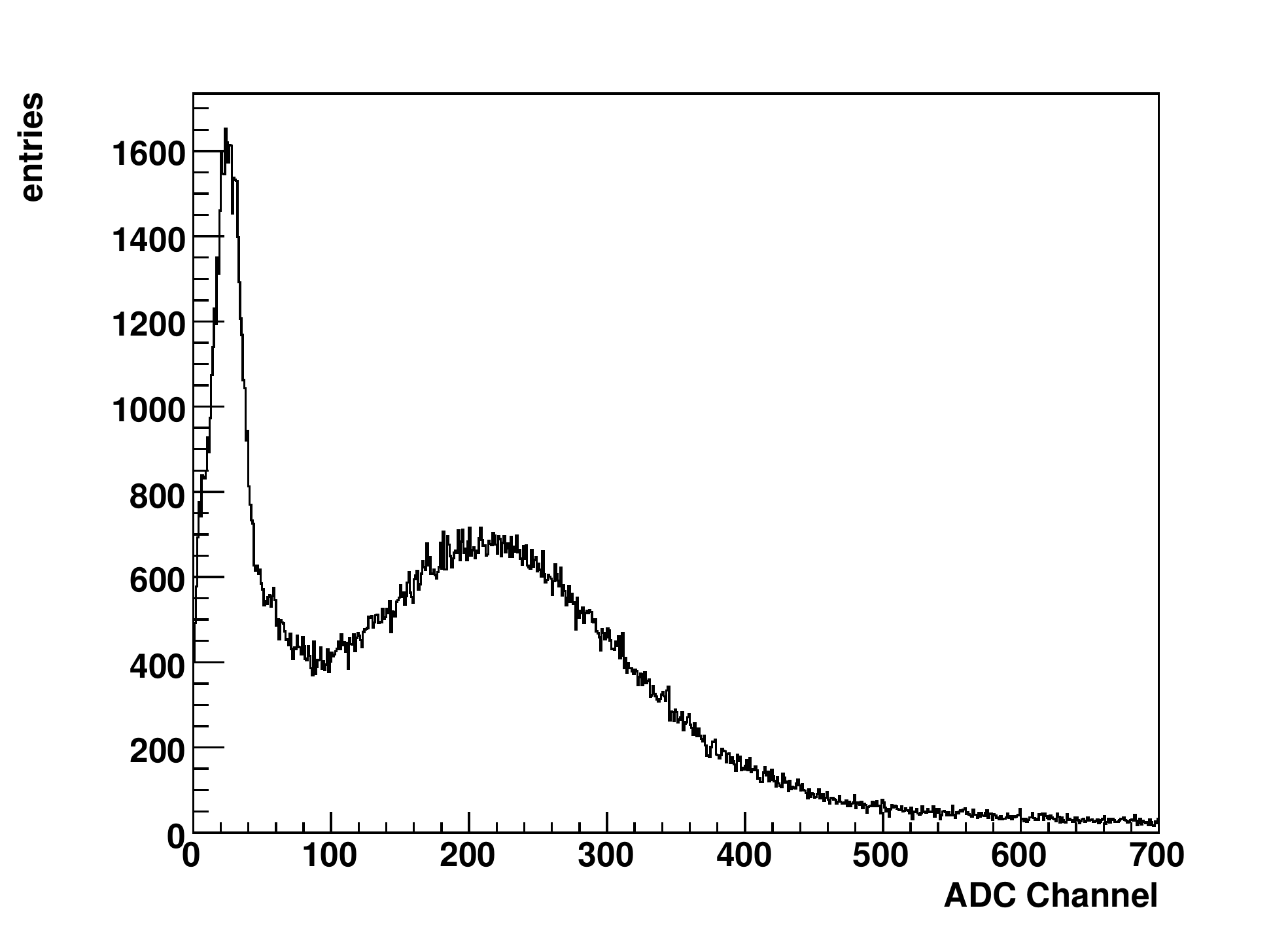}}}
   \end{tabular}
\end{center}
\caption{Pedestal-subtracted scintillation spectra in proton-irradiated
Cerium Fluoride, showing left the peak due to cosmic muons
(taken with 26 db attenuation), and the the right a peak possibly due to $^{139}$Ce
(0 db attenuation).\label{f-muon}}
\end{figure}

Such muons are minimum-ionising, and leave, according to
~\cite{r-dedx}, an energy deposit of 7.9 MeV/cm in the crystal.  Due to
the geometrical acceptance of the trigger scintillator setup, the mean
path in the crystal is $2.3 \pm 0.2$ cm. and thus the muon energy deposit in
average $18.2\pm 1.6$ MeV.  The Light Output spectrum is shown in
Fig.~\ref{f-muon} (top), which was acquired attenuating the
photomultiplier signal by 26 db.
The peak position is determined to be at channel $1276\pm 8$.  Keeping
the same PM gain, but with 0 db attenuation, we have acquired the
spectrum in Fig.~\ref{f-muon} (bottom), by triggering on the Light
Output from the crystal itself with a threshold set below the level of single
photoelectrons thermally emitted by the photocathode: those yield the
leftmost peak in the histogram. The peak at channel $225 \pm 1$
corresponds, if we determine its equivalent energy deposit scaling
from the muon peak position, to $E_{\gamma} = 160 \pm 10$ keV.  As
will be evident from activation measurements and related FLUKA
simulations described in the following sections, the dominant isotope
created in the proton irradiation tests is $^{139}$Ce, whose electron
capture decay to $^{139}$La is accompanied by an emission of a 165 keV
photon. The peak  we observe in the spectrum is
in good agreement with the activity from this isotope.

A Light Output measurement using cosmic muons one year after the
second irradiation allows us to determine
 a remaining loss of $ \Delta{\mathrm{LO / LO}} =
(11\pm 2) \%$.  The measured fraction of induced absorption
coefficient which has not recovered 1 year after the second
irradiation (see Fig.~\ref{f-mu}), is $\mu_{IND}=0.33 \pm 0.04 \;\mbox{m}^{-1}$.  A
correlation between LO loss and induced absorption has been published
in \cite{r-LYNIM} for 23 cm long Lead Tungstate crystals, and the
correlation therein between LO loss and induced absorption coefficients
is similar to the one we observe here. A precise
comparison would require taking into account the different crystal
dimensions and their influence on light collection. Our measurement
however shows how the observed spontaneous recovery at room
temperature of transmission loss after hadron irradiation in Cerium
Fluoride up to $\Phi_p=(2.12 \pm 0.15) \times
10^{14}\;\mathrm{cm^{-2}}$ is accompanied, as expected, by an almost
complete recovery of scintillation Light Output.

\section{Activation measurements and results}
\label{s-ACT}
\subsection{Present irradiations}
\begin{figure}[h]
\begin{center}
  \begin{tabular}[h]{cc}
{\mbox{\includegraphics[width=55mm,angle=90]{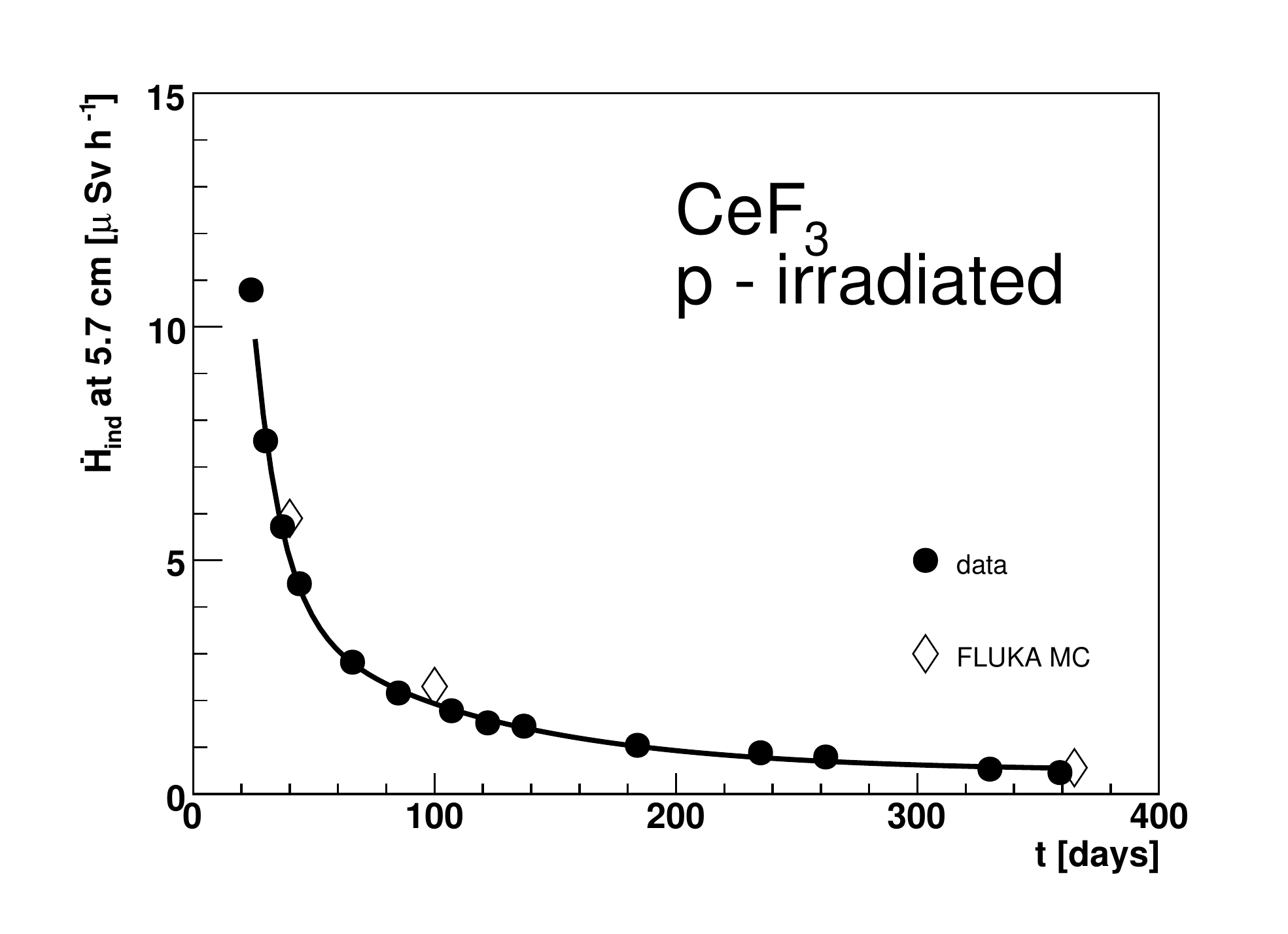}}}&
{\mbox{\includegraphics[width=55mm,angle=90]{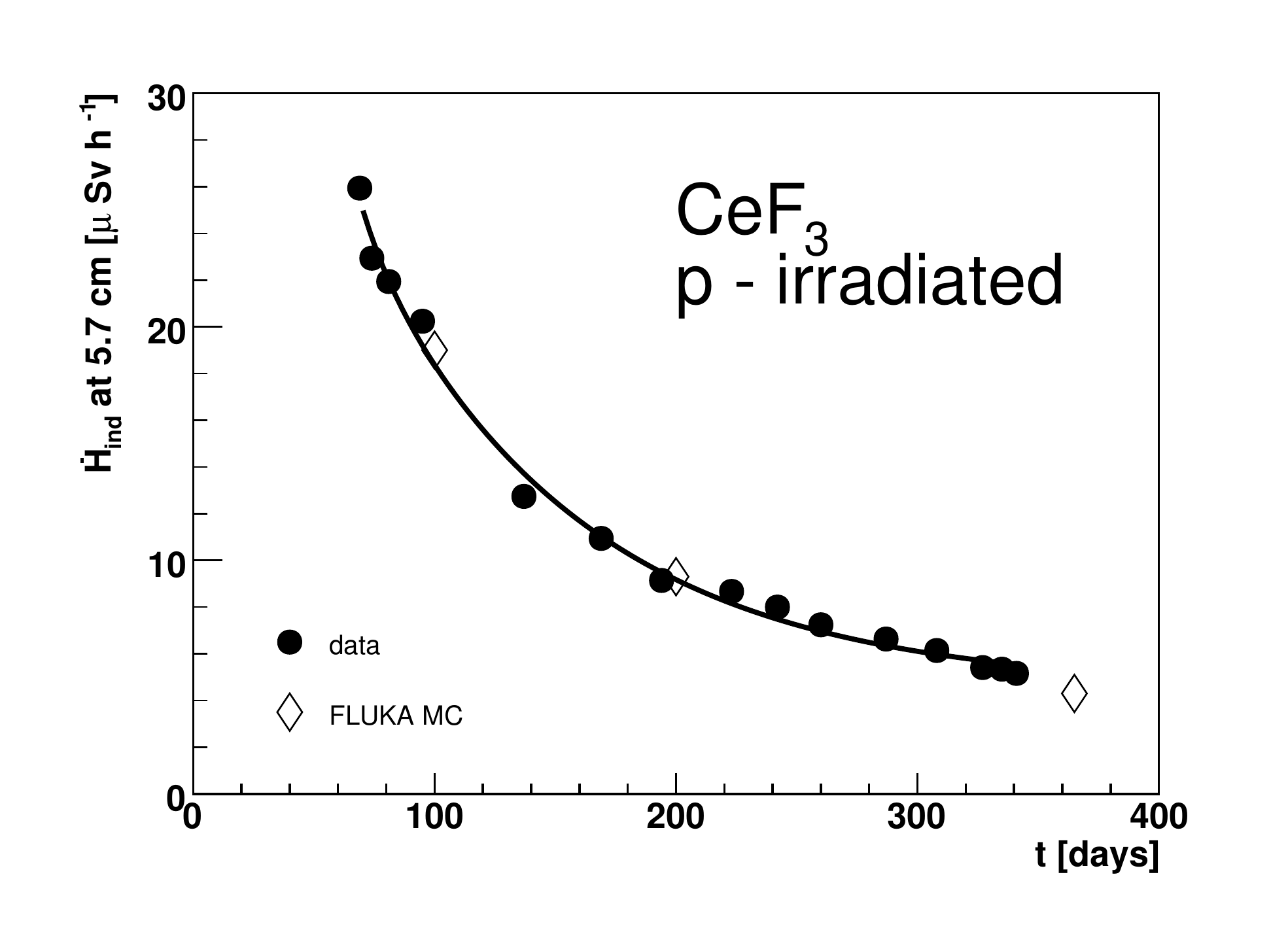}}}
   \end{tabular}
\end{center}
\caption{Measured remnant dose (black dots) as a function of cooling
time for Cerium Fluoride after the first
irradiation (left) and after the second irradiation (right) compared to
the expected values from FLUKA simulations (white lozenges).}\label{f-RA}
\end{figure}

Hadron irradiation causes the production of radioactive isotopes in the
crystals. While most of them are short-lived, those with a long
half-life are responsible for the remnant radioactivity and are
relevant in case a calorimeter needs human intervention after
exposure.  It might thus be of interest to compare measurements of
radio-activation in Cerium Fluoride to those in Lead Tungstate. The
latter has been extensively studied through simulations and
measurements in our early work~\cite{r-LTNIM} and references
therein. The measurements there agree with simulation results on
average within 30\% and never beyond a factor of 2, and confirm that
radiation exposure is an important concern for a Lead Tungstate
calorimeter used in intense hadron fluences.  Activation measurements
in Cerium Fluoride provide important practical information on access
and handling possibilities for such a calorimeter if used at superLHC.
The induced ambient dose equivalent rate (``dose'') ${\dot{H}^*(10)_{\rm ind}}$ was
regularly measured according to the procedure described in
\cite{r-LTNIM} with an Automess 6150AD6~\cite{r-AUTOMESS} at a
distance of 4.5 cm from the long face of the crystal at its
longitudinal centre. The reference point of the sensitive element in
the 6150AD6 is reported to be 12mm behind the entrance window. Thus
our actual distance was 5.7 cm from the crystal face.  The measured
dose as a function of cooling time is plotted in Fig.~\ref{f-RA}.
The activation values are compatible with the scaling of fluences,
if one takes into account that when the second irradiation was
started, the crystal was still showing a remaining level of activation
from the first one.
\begin{table}[hb]
\begin{center}
\begin{tabular}{| l |c|c|c|}
\hline
 Isotope & $\tau_{\frac{1}{2}} $ &Activity& total $\gamma$ \\
               &                                     & [Bq/cm~$^3$]& energy \\
\hline
$^3H$ & 12.33 y & $1017 \pm 3$ & - \\
$^{88}Y$ & 106.65 d & $109 \pm 1$ & 2.734 MeV \\
$ ^{109}Cd$ & 462.6 d  & $114 \pm 1$ & 0.088 MeV \\
$^{139}Ce$ & 137.64 d & $930 \pm 6$ & 0.166 MeV \\
\hline    
\end{tabular}
\end{center}
\caption{Dominating isotope activities in Cerium Fluoride from FLUKA
  simulations, 1 year after irradiation
  by 24 GeV/c protons up to a fluence
  $\Phi_p=2.78\times10^{13}$\,cm$^{-2}$ }
\label{t-ACT}
\end{table}  

We have fitted
the data, taken over one year, with a sum of a constant and two
exponentials with time constants $\tau_i\; (i=1,2)$:
\begin{equation}
\dot{H}^*(10)_{\rm ind}(t_{\rm{rec}}) = \sum_{i=1}^{2} D_i^je^{-t_{\rm{rec}}/\tau_i} + D_3^j,
\label{e-Dfit}
\end{equation}
where $t_{\rm{rec}}$ is the time elapsed since the irradiation, while
$D_i^j,\; (i=1,2)$ and $D_3^j$ are the amplitude fit parameters for
irradiation $j\; (j=1,2)$.  The time constants obtained for the two
independent fits are compatible, with values $\tau_1 = 11$ days and
$\tau_2 = 85$ days.
Interestingly, the radio-activation recovery time constants are
compatible within $2\sigma$ with those for the damage recovery
determined in Sec.~\ref{s-LT}.  While there is
no evidence for a link between the two, we wonder whether the
long-lived induced absorption component might be due to a
self-irradiation of the crystal.

FLUKA Monte Carlo simulations were performed using the code Version 2008.3c.0~\cite{r-fluka1, r-fluka2} and rely on the input parameters used in our $\mathrm{PbWO}_4$ study~\cite{r-HUH,r-LTNIM}. The beam profile was assumed to be squared ($3\times3\,\mathrm{cm^{2}}$) and uniformly distributed. The crystals was simulated according to the experimental setup. The FLUKA geometry includes also the back wall of the irradiation zone, i.e. the T7 beam line dump. Because the hadron shower induced from beam protons impinging on the crystal shows a significant forward direction, such that the integrated hadron fluence at the backside is roughly ten times more intense than the lateral fluence, the side walls were neglected.

For the dose measurements following the irradiation, the crystal has been removed from the irradiation zone and was kept for measurements in an area with low background. To simulate the two processes with different geometries, the FLUKA two-step method~\cite{r-2step} was applied. In a first step (the irradiation) the produced radionuclides in the crystal, namely the $\gamma$ - and $\beta^+$ - emitters, were recorded. These provided the input for the second step, where the average ambient dose equivalent in a volume of $1\,\mathrm{cm^{3}}$ was calculated, using fluence to ambient dose equivalent conversion coefficients~\cite{r-AMB74}. The center of the dose recording region was set laterally centered and at a distance of $5.7\,\mathrm{cm}$ from the crystal, according to the experimental settings.
The FLUKA simulation results of the Cerium Fluoride activation at a few
intervals after each of the two irradiations are shown in
Fig.~\ref{f-RA}.  One approximation was made in the simulation: a
single irradiation was assumed in each case, while the same crystal
was actually irradiated twice at one year's interval. This might
explain the slight dose underestimate from FLUKA at long
cooling times after the second irradiation.  Table~\ref{t-ACT} lists
the isotopes expected to be still present in the crystal, according to
FLUKA, one year after the irradiation up to a fluence
$\Phi_p=2.78\times10^{13}$\,cm$^{-2}$ and with an activity
larger than 10 Bq/cm$^3$.
From the tabulated values, it is evident that the isotope that
contributes with the largest activity is $^{139}$Ce.  One may also notice
that the longer recovery time constant fitted in Fig.~\ref{f-RA} for the radiation dose
is of the same order of magnitude as the life time of $^{139}$Ce.
Taking into account contributions from the other long-lived isotopes
present, the agreement is quite reasonable.

\subsection{Full-size crystals}
\begin{figure}[b]
\begin{center}
{\mbox{\includegraphics[height=11cm]{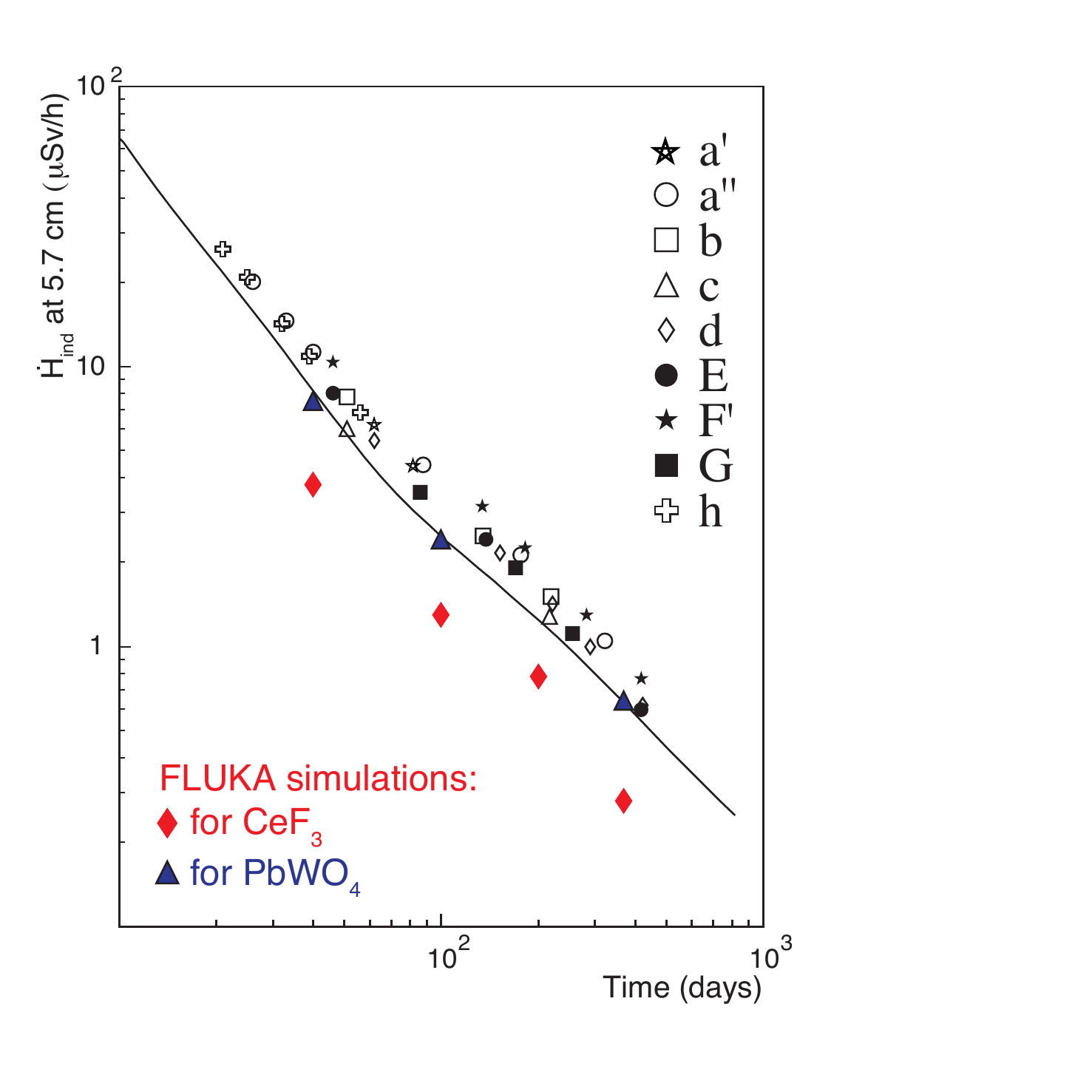}}}
\end{center}
\caption{FLUKA simulation of a 42 cm long $\mbox{CeF}_3$ crystal
  remnant dose as a function of cooling time (full lozenges),
  compared to data (symbols) and simulation (line) results on 23 cm
  long Lead Tungstate crystals from Ref.~\cite{r-LTNIM} for a fluence
  $\Phi_p$=$10^{13}$\,cm$^{-2}$ of 20 GeV/c protons.
  Full triangles indicate, for
  validation purposes, our simulation results for Lead Tungstate.}
\label{f-decay}
\end{figure}
A comparison between Cerium Fluoride and Lead Tungstate activation
levels after hadron irradiation for full-size crystals is a relevant
input to the selection of the calorimetric medium for a calorimeter upgrade.
In Fig.~\ref{f-decay} we show the results from our FLUKA simulation for
the irradiation and cooling conditions of \cite{r-LTNIM} for full-size
Lead Tungstate crystals exposed to 20 GeV/c protons,
compared to the measured activation and the
FLUKA simulation performed therein. 
 The agreement validates the
present FLUKA simulations.  Also shown in Fig.~\ref{f-decay} are the
FLUKA results for the activation expected for full-size Cerium
Fluoride crystals.  Because of the smaller density and longer radiation
length of Cerium Fluoride, a full-size calorimeter crystal would have to be 42
cm in length. Transverse dimensions of 24 mm x 24 mm correspond to
the typical cross section yielding a similar $\eta - \phi$ granularity
as for existing Lead Tungstate calorimeters.  The FLUKA simulation, which was
validated already for short Cerium Fluoride crystals (Fig.~\ref{f-RA})
and for Lead Tungstate (Fig.~\ref{f-decay}), was performed
for such full-size dimensions and for the same fluence of $\Phi_p=1
\times 10^{13}\;\mathrm{cm^{-2}}$, yielding dose values shown as lozenges
in Fig.~\ref{f-decay}.  One observes that workers' exposure to 
proton-irradiated, 26 X$_0$ long Cerium Fluoride crystals is expected
to be always a factor of 2 to 3 lower than the one due to Lead Tungstate.

\section{Conclusions}
\label{s-CON}
We have studied Cerium Fluoride as a possible scintillating crystal
for calorimetry at the superLHC. This investigation was inspired by
our earlier studies of Lead Tungstate, where we observed a
hadron-specific, cumulative damage from charged hadrons. 
All characteristics of the damage in Lead Tungstate are consistent 
with an intense local energy deposition from heavy
fragments.  Measurements of absorption induced in $\mathrm{CeF}_3$ by
24 GeV/c protons up to fluences $\Phi_p=(2.78 \pm 0.20) \times
10^{13}\;\mathrm{cm^{-2}}$ and $\Phi_p=(2.12 \pm 0.15) \times
10^{14}\;\mathrm{cm^{-2}}$ show a Light Transmission
damage which is not cumulative, is
more than one order of magnitude smaller than in $\mathrm{PbWO}_4$
6 months after irradiation, and --- unlike
$\mathrm{PbWO}_4$ --- recovers further.  The absence of a
dominant Rayleigh-scattering component in $\mathrm{CeF}_3$ confirms
our understanding, that in $\mathrm{PbWO}_4$ it is due to
highly-ionising fission fragments as produced in crystals with
elements above Z=71. The scintillation Light Output in  $\mathrm{CeF}_3$
is observed to recover by 90\% over 1 year, and the remaining loss is consistent
with the induced absorption still present.

With its extreme resistance to hadron-induced damage, manifested
through a modest induced absorption which recovers with time, low
raw material costs, high light yield and negligible temperature
dependence, Cerium Fluoride is an excellent candidate for medical
imaging applications and for calorimetry at superLHC or in any high hadron
fluence environment.

\section*{Acknowledgements}
We are indebted to R.~Steerenberg, who provided us with the required
CERN PS beam conditions for the proton irradiations. We are deeply
grateful to M.~Glaser, who operated the proton irradiation facility
and provided the Aluminium foil dosimetry.

\end{document}